%% file: main.tex
\documentclass[10pt, conference, letterpaper]{IEEEtran}

\usepackage[english]{babel}
\usepackage{blindtext}
\usepackage{booktabs}
 
\usepackage[T1]{fontenc}
\usepackage{graphicx}

\usepackage{subfigure}
\usepackage{enumitem}
\usepackage{color}
\usepackage{url}
\usepackage[font=small]{caption}

\usepackage{algorithm}
\usepackage{algorithmicx}
\usepackage{algpseudocode}

\newcommand{\dbsub}[1]{#1}
\renewcommand{\dbsub}[1]{\textit{<blanked>}}

\newcommand{\myitem}[1]{\vspace{0.25\baselineskip}\noindent\textbf{#1}}

\newcommand{\rv}{RouteViews\xspace}

\newcommand{\perc}{\,\%\xspace}

\newcommand{\etal}{et al.\xspace}
\newcommand{\one}{(i)~}
\newcommand{\two}{(ii)~}
\newcommand{\three}{(iii)~}

\newcommand{\pamout}[1]{}
\newcommand{\pam}[1]{#1}
\newcommand{\tmaout}[1]{}
\newcommand{\tma}[1]{#1}
\newcommand{\camr}[1]{#1}

\newcommand{\imp}{IMP\xspace}
\newcommand{\imps}{IMPs\xspace}
\newcommand{\vp}{VP\xspace}
\newcommand{\vps}{VPs\xspace}
\newcommand{\mathvp}{\mathcal{V}}

\usepackage{cleveref}
\newcommand{\secref}[1]{\S\ref{#1}}
\usepackage{cite}

\author{
\IEEEauthorblockN{Pavlos Sermpezis\textsuperscript{$\dagger$}, %
Lars Prehn\textsuperscript{$\star$}, %
Sofia Kostoglou\textsuperscript{$\dagger$}, %
Marcel Flores\textsuperscript{$\ddagger$}, %
Athena Vakali\textsuperscript{$\dagger$}, %
Emile Aben\textsuperscript{$\mathparagraph$}%
}
\vspace{0.7\baselineskip}
\IEEEauthorblockA{
\textsuperscript{$\dagger$} Aristotle University of Thessaloniki; \{sermpezis, sofikost, avakali\}@csd.auth.gr
\\
\textsuperscript{$\star$}Max Planck Institute for Informatics; lprehn@mpi-inf.mpg.de
\\
\textsuperscript{$\ddagger$}Edgio; mflores@edg.io
\\
\textsuperscript{$\mathparagraph$}RIPE NCC; emile.aben@ripe.net
}
}

\IEEEoverridecommandlockouts

\begin{document}
\title{
Bias in Internet Measurement Platforms
}

%



\maketitle

\begin{abstract}
Network operators and researchers frequently use Internet measurement platforms (IMPs), such as RIPE Atlas, RIPE RIS, or RouteViews for, e.g., monitoring network performance, detecting routing events, topology discovery, or route optimization. To interpret the results of their measurements and avoid pitfalls or wrong generalizations, users must understand a platform's limitations. To this end, this paper studies an important limitation of IMPs, the \textit{bias}, which exists due to the non-uniform deployment of the vantage points. Specifically, we introduce a generic framework to systematically and comprehensively quantify the multi-dimensional (e.g., across location, topology, network types, etc.) biases of IMPs. Using the framework and open datasets, we perform a detailed analysis of biases in IMPs that confirms well-known (to the domain experts) biases and sheds light on less-known or unexplored biases. To facilitate IMP users to obtain awareness of and explore bias in their measurements, as well as further research and analyses (e.g., methods for mitigating bias), we publicly share our code and data, and provide online tools (API, Web app, etc.) that calculate and visualize the bias in measurement setups.
\end{abstract}

\section{Introduction}\label{sec:intro}
\input{sections/intro}

\section{Internet Measurement Platforms and Bias: a Primer}\label{sec:preliminaries}
\input{sections/preliminaries}

\section{\tma{Quantifying Bias: Framework \& Methodology}}
\label{sec:methodology}
\input{sections/methodology}

\section{Analyzing \imp bias}\label{sec:bias-results}
\input{sections/analysis}



\section{Open Data, Code,  API, and Tools}\label{sec:portal}
\input{sections/portal}

\section{Related Work}\label{sec:related}
\input{sections/related}

\section{Conclusion}\label{sec:conclusion}
\input{sections/conclusion}

\section*{Acknowledgements}
This research is co-financed by Greece and European Union through the Operational Program Competitiveness, Entrepreneurship and Innovation under the call RESEARCH-CREATE-INNOVATE (project T2EDK-04937) and RIPE NCC (AI4NetMon project).

\begin{figure*}[th!]
    \centering
    \subfigure[KL divergence]{\includegraphics[width=0.3\linewidth]{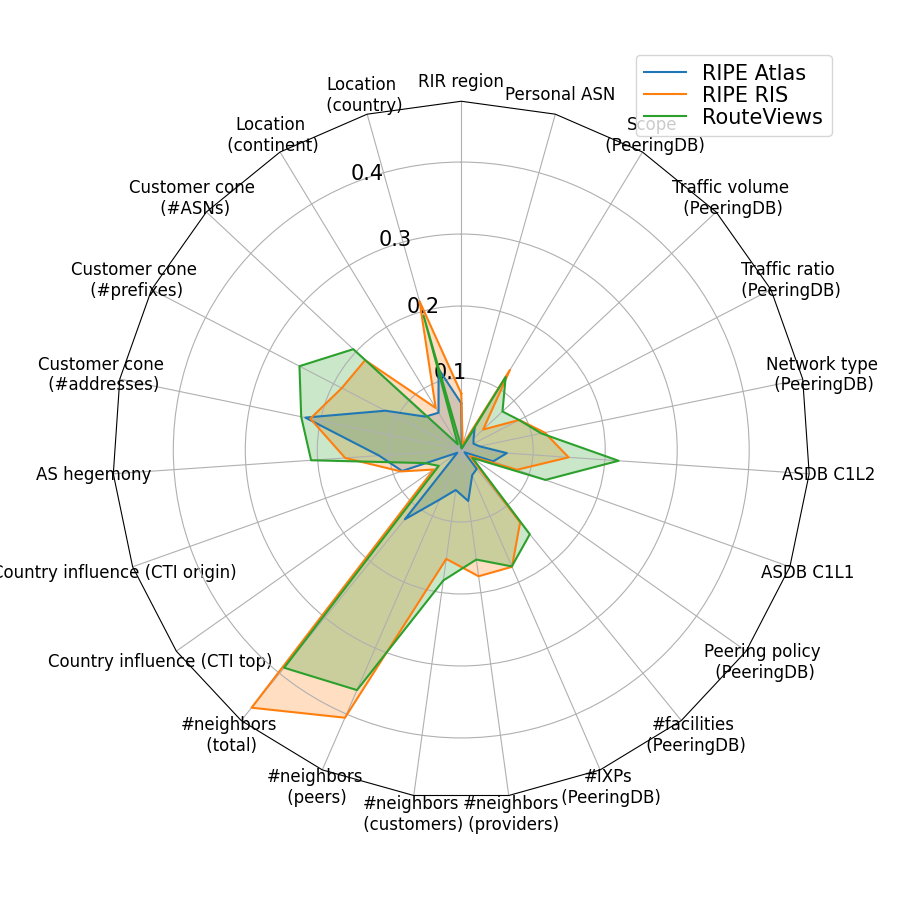}\label{fig:bias-vs-metrics-kl}}
    \subfigure[Total variation]{\includegraphics[width=0.3\linewidth]{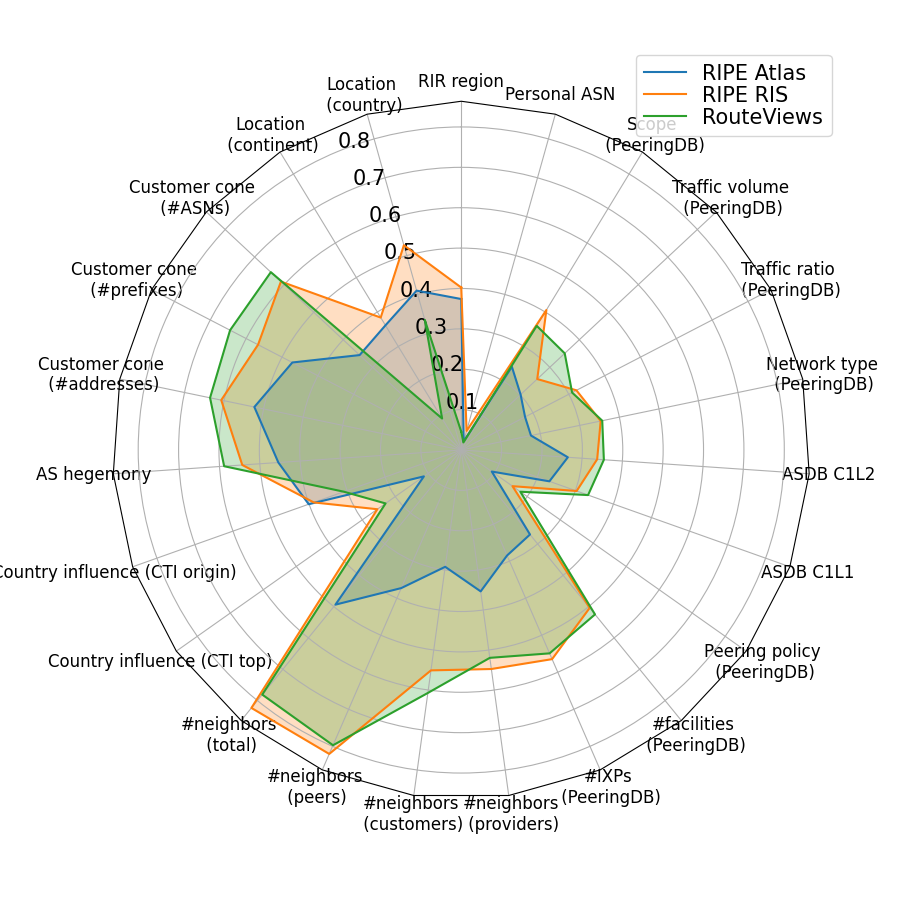}\label{fig:bias-vs-metrics-tv}}
    \subfigure[Max distance]{\includegraphics[width=0.3\linewidth]{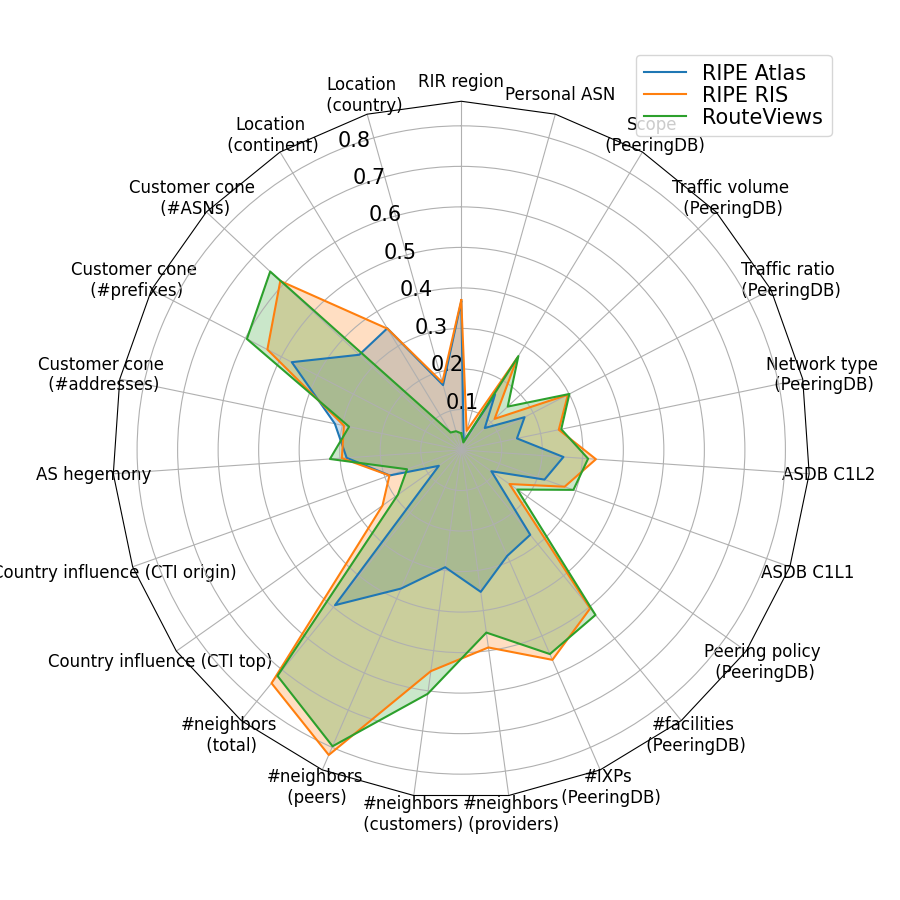}\label{fig:bias-vs-metrics-max}}
    \caption{Radar plot depicting the bias score for different bias metrics.}
    \label{fig:bias-vs-metrics}
\end{figure*}

\bibliographystyle{splncs04}
\bibliography{reference_short}

\appendices


\input{appendix}


\end{document}

%% file: sections/intro.tex
Public Internet Measurement Platforms (IMPs) like RIPE Atlas~\cite{ripe-atlas}, RIPE RIS~\cite{ripe-ris}, or RouteViews~\cite{routeviews} are fundamental building blocks of networking research and operations. Network operators and researchers frequently use their measurement capabilities and publicly archived data to, e.g., detect routing events and malicious networks~\cite{sermpezis2018artemis,fontugne2019bgp,testart2019profiling}, analyze the Internet's structure~\cite{arnold2020cloud,oliveira2008search,gregori2012incompleteness}, understand and optimize (their own) routing policies~\cite{sermpezis2019inferring,streibelt2018bgp,gray2020bgp}, or detect outages and performance bottlenecks~\cite{ra2017seeing,giotsas2017detecting,shah2017disco}.

IMPs operate a broad range of globally distributed vantage points. 
While RIPE Atlas hosts around 11,000 measurement probes in 3,300 autonomous systems (ASes), RIPE RIS and \rv collect routing information from around 300 and 500 ASes, respectively. Despite their presence in thousands of ASes, IMPs only provide a limited view into the routing ecosystem. It is well-known that \imps capture incomplete views of the Internet~\cite{arnold2020cloud,ager2012anatomy,prehn2022peering,giotsas2013inferring,oliveira2008search} and sometimes offer misleading or incomplete answers for seemingly simple questions~\cite{del2019filtering,roughan201110,holterbach2015quantifying,willinger2009mathematics}. This incompleteness problem resulted in approaches for extending the observed AS topology via other data sources~\cite{chen2009sidewalk,faggiani2014study,arnold2020cloud,battista2006extract,giotsas2013inferring} or by adding new, favorably-positioned vantage points to \imps~\cite{roughan2008bigfoot,gregori2012incompleteness,leyba2022cutting,cittadini2014quality}.
\tma{While deploying \imp infrastructure aiming to increase completeness (e.g., "hunting for the most AS links" by deploying route collectors to IXPs) has a clear value, it frequently leads to \textit{unequal} (or, "biased") visibility of different parts of the Internet. This bias can come along many dimensions, such as network types, geographic placement, etc.  
} 

\tma{Despite extensive studies on \textit{incompleteness}, it remains unclear how \textit{representative} our view of the entire Internet routing ecosystem is: \textit{Do we have equal visibility to all types of networks? And, if not, how biased are the views we obtain from IMPs?}}
\tma{
In this paper, we study this unexplored aspect and take first steps towards a comprehensive characterization of the bias in \imps. Contrary to previous works that focus on specific aspects of bias, we argue that capturing representativeness is an inherently multi-dimensional problem. 
To this end, we make the following contributions:
}





\begin{itemize}[leftmargin=*,nosep]
    \item We formally define bias, and introduce a generic framework for quantifying the bias in a multi-dimensional context (\secref{sec:bias-definition}). \tma{The framework receives as input information about characteristics of networks (connectivity, location, etc.), and quantifies how representative a set of vantage points is.}
    
           
    \item \tma{We aggregate information from real-world datasets (\secref{sec:bias-dimensions}), and apply our framework to quantify the bias of widely-used \imps 
    (\secref{sec:bias-results}).}
    Despite the numerous limitations that come with real-world data sets (e.g., abstractions, inaccuracies, or incompleteness), we observe that our framework is capable of replicating the findings of previous studies (e.g., \cite{roughan201110,bajpai2015lessons,bajpai2017vantage}) conducted by domain experts. Besides these well-known issues, our framework can produce novel, and more nuanced, insights about the bias in \imps.
    
    \indent \tma{For example, our analysis confirms that RIPE RIS is heavily biased towards larger networks and IXPs~\cite{roughan201110}, and it also reveals that while networks that peer at many IXPs are over-represented in RIPE RIS, their peering policies (PeeringDB) are representative of the Internet's peering ecosystem.}
    


    \item \tma{We extend our analysis to explore the improvement potential of \imps (\secref{sec:improvement-potential}), and study the biases involved in common measurement practices, such as RIPE Atlas probes selection or use of individual route collectors (\secref{sec:bias-common-measurements}).}

    \item \tma{We publicly share our code and data~\cite{ai4netmon-github}, and discuss how it can be parametrized to extend or adapt our analyses. Moreover, to further facilitate users to explore and quantify bias in \imps or in custom measurement setups, we provide an API and a web portal with interactive visualizations (\secref{sec:portal}).}
    
    \tmaout{
    \item Leveraging our framework, we can design methodologies to reduce bias in existing \imps (\secref{sec:sampling-methods}). Carefully selecting subsets of \imp vantage points ("subsampling") can lead to significantly more representative measurements. We demonstrate this through a use case, in which we estimate the client latency distribution of a large content delivery network (CDN) through RIPE Atlas measurements: subsampling can lead to a more accrurate estimation than randomly selecting probes, or even than using the entire set of Atlas probes.
    }
    
    \tmaout{
    \item Another way for reducing bias is by extending the current \imp infrastructure (\secref{sec:extra-infrastructure}). Using RIPE RIS as an example, we calculate the bias difference each AS would introduce upon connecting to the platform. Since not every AS may be equally easy to acquire as participant in RIS, we collect data from domain experts and try to infer the acquisition complexity per AS. Our findings show that there are many easy-to-peer-with ASes that would reduce RIPE RIS bias. 
    }



\end{itemize}

\noindent \tma{We believe that having a framework to 
systematically quantify bias (and tools that automate it) can be valuable for Internet measurements: e.g., from \textit{raising awareness} to users (and lowering the bar for domain expertise) so that they avoid pitfalls in interpretation of measurement data, to generating a "\textit{bias assessment}" for each measurement study.}


\tma{We deem our work as a first step in this direction. In \secref{sec:conclusion}, we provide a critical discussion about how our analysis can be extended or refined to overcome existing limitations, and future research directions that can build upon our framework.}

%% file: sections/preliminaries.tex
In this section, we introduce the concept of bias on a general example (summarized in Table~\ref{table:toy-example}). Afterwards, we introduce the three major \imps that we analyze in this paper (\secref{sec:infrastructure}), discuss some of their known biases, and motivate the research questions that our study aims to address (\secref{sec:bias-examples-infra}). 

Let us assume a population consisting of 100 people, 50 of which are men and 50 women. If we run a survey with 10 people, of which 8 men and 2 women, our sample is biased towards men. We say that our sample is biased as there is a \textit{difference in the distributions between the entire population and our sample}.

\begin{table}[h]
    \centering
    \caption{Bias example: population and sample statistics. The gender bias is 0.22  (KL-divergence metric; \secref{sec:bias-definition}) and is higher than the country bias 0.03.}
    \begin{small}
    \begin{tabular}{l|cc|cc}
         {}&{Men}&{Women}&{Country A}&{Country B} \\
         \hline
         {Entire population}&{50\%}&{50\%}&{70\%}&{30\%}\\
         {Survey sample}&{80\%}&{20\%}&{80\%}&{20\%}
    \end{tabular}
    \end{small}
    \label{table:toy-example}
\end{table}

\myitem{Measuring bias.} To \textit{identify} this bias, one could run statistical tests (e.g.,  Kolmogorov-Smirnov test) to compare the two distributions. To further \textit{quantify} the bias, it is common to measure the \textit{distribution distance} among the population and the sample distributions (e.g., with the Kullback-Leibler divergence metric)

\myitem{Multi-dimensional bias.} Let us consider that our survey focuses on the height of individuals. If we compare the distributions of height within our total population to that within our survey sample, we may find that they differ as men (who naturally tend to be around \textasciitilde7\perc taller~\cite{World2022Height}) are over-represented. Now, let us consider that our survey further focuses on the native language of individuals. For this second case, the gender-bias in our sample would not affect our findings. In contrast, the country-bias (e.g., see the right side of the Table~\ref{table:toy-example}) of our sample, may play a major role. In other words, 
\textit{different bias dimensions (e.g., gender or country) may affect our measurements findings differently, depending on how they relate to the insights we want to gain.}

\subsection{\imps: RIPE Atlas, RIPE RIS, \rv}\label{sec:infrastructure}
We briefly overview the 3 major \imps on which we focus. 

\myitem{RIPE Atlas~\cite{ripe-atlas}} is a platform that hosts more than 11,000 measurement "probes" in more than 3,000 ASes. \pamout{While probes initially started out as solely physical devices, nowadays many probes are so called software probes, i.e., they run as apps on general-purpose devices. }Probes support a fixed set of measurement types (e.g., ping, traceroute, DNS)
. Users can select sets of probes and execute measurements (e.g., a traceroute towards a target IP), under some \pamout{per-probe and per-user }rate-limits.

\myitem{RIPE RIS~\cite{ripe-ris} and \rv~\cite{routeviews}} are two global platforms that 
host "route collectors", which are dedicated devices that passively receive, dump, and publicly archive the routing information from their peering networks. Most route collectors are located at large IXPs such that they can quickly establish many sessions over the IXP's peering LAN. The "multi-hop"-enabled route collectors may establish indirect sessions with remote ASNs. In total, RIPE RIS and \rv host 27 (of which 3 multi-hop) and 36 (20 multi-hop) route collectors with more than 500 and 300 peer ASNs, respectively. A peering ASN may provide feeds for the entire routing table ("full feed") or only a part of it.

\subsection{\tma{\imp biases: known and unknown aspects \& user awareness}}
\label{sec:bias-examples-infra}


\begin{figure}
    \centering
    \subfigure[Location]{\includegraphics[width=0.75\linewidth]{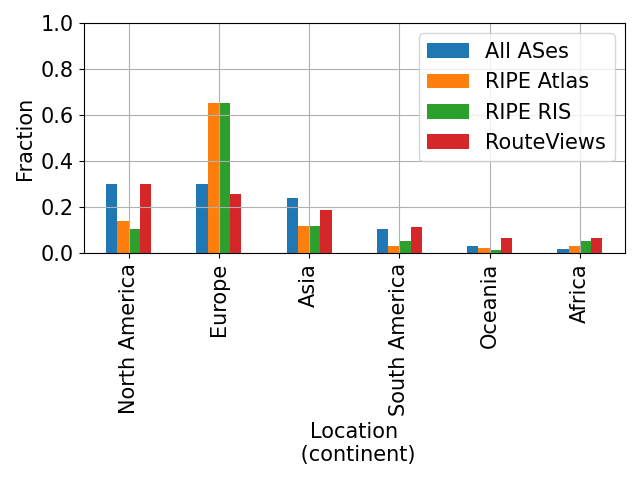}\label{fig:example-distribution-location}}
    \subfigure[Network type]{\includegraphics[width=0.75\linewidth]{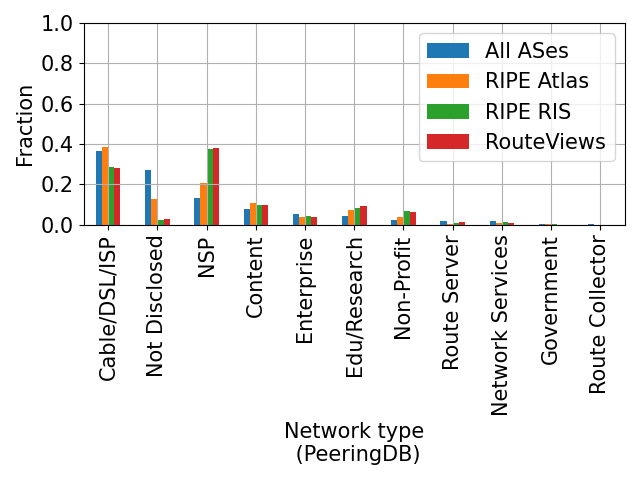}\label{fig:example-distribution-type}}
    \caption{Distributions of (a) \textit{Location (continent)} and (b) \textit{Network type} for the entire population of ASes (blue bars) and the set of ASes that are part of the RIPE Atlas, RIPE RIS, and RouteViews platforms.}
    \label{fig:example-distribution-characteristics}
\end{figure}
\myitem{Location bias.} A glance at the map with the locations of RIPE’s infrastructure (see~\cite{atlas-map} for Atlas probes and~\cite{ris-map} for RIS route collectors) reveals a higher density of the infrastructure in Europe, which is in imbalance with the spread of ASes around the world, i.e., there is location bias in RIPE Atlas and RIPE RIS. \tma{While this bias is well-known (or, easy to spot), a clear quantification is missing: \textit{how far are we from an ideal scenario? Or, what is the room for improvement?}}


On the contrary, Fig.~\ref{fig:example-distribution-location} shows that \rv has route collectors deployed in more representative locations around the world. \tma{However, \textit{are users of \rv and RIS aware of this significant difference (and do they take it into account in their measurements)? And, can combining measurements from both platforms significantly reduce the location bias?}}


\myitem{Topological bias.} Route collectors (RIPE RIS and RouteViews) are biased towards larger core networks and at Internet eXchange Points (IXPs) as reported in previous work~\cite{roughan201110}. \tma{\textit{How could we enable novice users (without "10 years" of experience~\cite{roughan201110}) to easily identify such biases?}}

\myitem{Bias awareness.} Neither the location nor the topological bias are new to expert users. However, not even expert-users might be able to accurately judge the extent of different biases on different \imps\tmaout{, e.g., while RIS and Atlas have substantial location bias, this bias is almost negligible for RouteViews}. Similarly, other biases along (less prominent) dimensions, such as the network type (see Fig.~\ref{fig:example-distribution-type}), might be even harder to judge. A questionnaire related to the topic of this paper that we ran
{ (see details in Appendix~\ref{appendix:survey})} supports the fact that not all users are aware of biases: out of the 50 questioned operators and researchers, only 26 (52\%) consider \imps to be biased, while 28\% consider that there is no bias (or, probably not), and 20\% "do not know". \pam{\textit{This lack of (or, partial) awareness motivates our study to comprehensively quantify the bias in \imps.}}





%% file: sections/methodology.tex
Similarly to the example of~\secref{sec:preliminaries} where people are characterized by two features (gender and origin country), the \imps can also be characterized by a multitude of features, such as location, connectivity, traffic levels, etc.. Each characteristic/feature can be considered as a dimension, and the bias can be calculated over each dimension. Then, depending on the measurement use case, all or some of the dimensions can be taken into account, depending on their relevance (see~\secref{sec:preliminaries}). 

\tma{In this section, we first introduce our framework, and formally define the bias and the metrics to quantify it (\secref{sec:bias-definition}). The framework is generic: it takes as input a dataset of network characteristics, and returns in a unified way the bias along all characteristics (i.e., ``bias dimensions'').
Then, in \secref{sec:bias-dimensions}, we present the dataset we compile and use in this paper as input to our framework for analyzing the bias of \imps.
}


\subsection{\tma{The bias quantification framework}}
\label{sec:bias-definition}

\myitem{Definitions.} Let $P$ be the distribution of a characteristic (e.g., network size) within a set of networks $\mathcal{N}$. If the characteristic takes $K$ distinct values, its distribution is $P = [p_{1}, ..., p_{K}]$, where $p_{i}$ is the probability of a network having the $i$ value (e.g., $p_{Europe}$=0.32 for the entire population of ASes; see Fig.~\ref{fig:example-distribution-location}); formally, $p_{i}=\frac{1}{|\mathcal{N}|}\sum_{j\in\mathcal{N}}I_{j\rightarrow i}$, where $I_{j\rightarrow i}$ an indicator function that is 1 if the network $j$ has the characteristic $i$\pamout{ (and is 0 otherwise)}, and $|\mathcal{N}|$ the size of the set $\mathcal{N}$.

Also, let a subset of networks $\mathvp\subset\mathcal{N}$, and $Q$ be the corresponding distribution within the set of networks $\mathvp$. We define the bias (of the set $\mathvp$ wrt. to the set $\mathcal{N}$) as the distance between the distributions $P$ and $Q$.


\myitem{Identifying bias.} If the distance between $P$ and $Q$ is statistically significant, then there is bias. There are several statistical tests that could be applied. We use the Kolmogorov-Smirnov (or, KS-test), which is a nonparametric test that compares two distributions (two-sample KS-test)\pamout{\footnote{If $\mathcal{X}$ is the set of values that a characteristic can take, then the KS-test examines the maximum distance between the two (cumulative) distributions, i.e., $D = \max_{x\in\mathcal{X}} |P_{x}-Q_{x}|$.}}\tma{. The KS-test answers the question ``what is the probability that $P$ and $Q$ are drawn from the same distribution?''. If this probability is small enough (e.g., less than 5\%), then we can confidently state that the population (i.e., all ASes) and sample (i.e., \vps) follow different distributions, i.e., there is bias.
}

\myitem{Bias Metrics.} There exist several metrics to quantify the distance between two distributions. A metric that is commonly used (in particular, for concepts related to bias, e.g.,~\cite{steck2018calibrated}) is the Kullback–Leibler (KL) divergence:
\begin{equation}\label{eq:bias-metric-kl}
    B_{KL} = \textstyle \sum_{i=1}^{K}p_{i}\cdot \log \left(\frac{p_{i}}{q_{i}}\right)
\end{equation}
The KL-divergence takes on values in $[0, +\infty]$, where the higher the value the more the two distributions differ. In the paper, we use a bounded version of the KL-divergence that takes values in $[0,1]$~\cite{steck2018calibrated,Sacharidis2019ACA}\footnote{We substitute $q_{i}\rightarrow (1-w)\cdot q_{i} + w\cdot p_{i}$, with $w=0.01$ and normalize with its upper bound $\log\frac{1}{w}$, to get $B_{KL}=\frac{1}{\log\frac{1}{w}}\cdot \sum_{i\in\mathcal{K}}p_{i}\cdot \log \left(\frac{p_{i}}{(1-w)\cdot q_{i} + w\cdot p_{i}}\right)$.}\pam{, and we call it the \textit{bias score}}. \tma{We calculate a bias score per characteristic/dimension.}

For example, in terms of the location distributions depicted in Fig.~\ref{fig:example-distribution-location}, the bias score for RIPE Atlas and RIPE RIS is $B_{KL}=0.06$ and $B_{KL}=0.07$, respectively, while for RouteViews, which follows a similar distribution to the entire population, the bias score is $B_{KL}=0.01$. For the network type (Fig.~\ref{fig:example-distribution-type}) the bias scores for RIPE Atlas, RIPE RIS, and RouteViews are $0.03$, $0.12$, and $0.11$, respectively, clearly highlighting the higher bias in the route collector projects.

\textit{Remark:} \pam{We tested other common metrics (e.g,. Total Variation) for the bias score as well \tma{(see Appendix~\ref{appendix:bias-metrics})}}. While the actual values of each metric are different, the qualitative findings of the paper remain the same.  

\tma{The framework is generic with respect to $\mathcal{N}$, $\mathcal{V}$, bias dimensions, bias metrics, and input data. In this paper, we consider as 
\camr{
\begin{itemize}[leftmargin=*]
    \item $\mathcal{N}$: the entire population of ASes (i.e., more than 100,000 ASNs for which we have data)
    \item $\mathcal{V}$: the set of \vps of an \imp (e.g, the peers of RIPE RIS  or \rv, or the probes of RIPE Atlas)
\end{itemize}
}
\noindent and in the next section we compile a dataset of several network characteristics/dimensions. Later, in~\secref{sec:portal}, we discuss how other choices can be done for these parameters.}

\pamout{Another common metric is the Total Variation distance (TV), which is defined as $B_{TV} =  \frac{1}{2}\sum_{i=1}^{K}|p_{i}-q_{i}|$. TV takes also values in [0,1]. The main difference between the two metrics is that the KL-divergence is more sensitive to changes in characteristics of lower probabilities $p_{i}$~\cite{steck2018calibrated}. However, while the actual values of each metric may be different, our results show that our qualitative findings do not change when using the TV metric (see Appendix~\ref{appendix:bias-metrics}).}


\subsection{\tma{Data and bias dimensions}}
\label{sec:bias-dimensions}


\tma{
\myitem{Data sources.} We take into account the characteristics of the \imps at an AS-level granularity (e.g., two RIPE Atlas probes in the same AS have the same AS-level characteristics). The reasons for this choice is twofold: data availability and  scope. Specifically, at the AS-level there are several public datasets: at a finer granularity there is scarce information (which would limit our analysis to only a few dimensions) and compiling a rich dataset would need extensive measurements per dimensions (which could be done only per use case, and thus would be beyond the scope --and space limitations-- of this paper). Nevertheless, our framework is extensible to a more fine-grained level (e.g., per monitoring device, such as at a vantage point level or router level); we discuss these extensions and limitations of our analysis in \secref{sec:conclusion}. 
}

We compile a list of characteristics for each AS \tma{from the widely used CAIDA AS-rank~\cite{CAIDA2022ASRANK} and AS-relationships~\cite{CAIDA2022ASREL} and PeeringDB~\cite{PDB2022WEB,CAIDA2022PDB} datasets, as well as from public datasets that contain information for the network size/importance (Internet Health Report's AS-hegemony metric~\cite{IHR,fontugne2018thin}, and the Country-level Transit Influence index, or CTI,~\cite{gamero2022quantifying}) and network types (bgp.tools~\cite{BGPtools2022personal} and ASDB~\cite{ziv2021asdb})}.

\tmaout{
\myitem{Data sources.} We take into account the characteristics of the \imps at an AS-level granularity (e.g., two RIPE Atlas probes in the same AS have the same AS-level characteristics). Nevertheless, our methodology and analyses are extensible and applicable to a more fine-grained level (e.g., per monitoring device, such as at a vantage point level or router level; see also the discussions in \secref{sec:portal} and \secref{sec:conclusion}). We compile a list of characteristics for each ASes \tma{from the widely used CAIDA AS-rank~\cite{CAIDA2022ASRANK} and AS-relationships~\cite{CAIDA2022ASREL} and PeeringDB~\cite{PDB2022WEB,CAIDA2022PDB} datasets, as well as from public datasets that contain information for the network size/importance (Internet Health Report's AS-hegemony metric~\cite{IHR,fontugne2018thin}, and the Country-level Transit Influence index, or CTI,~\cite{gamero2022quantifying}) and network types (bgp.tools~\cite{BGPtools2022personal} and ASDB~\cite{ziv2021asdb})}.  

\textit{Remark:} \pam{Our choice for AS-level granularity is twofold: data availability and  scope. Specifically, at the AS-level there are several public datasets, while at at a finer granularity there is scarce information; this would limit the generality of our analysis in terms of bias dimensions. For a finer granularity, one would need to conduct measurements and analyses to collect or infer the needed data, which can be very useful for several use cases, but is out of the scope (and space limitations) of this paper that aims to give a first comprehensive characterization of bias in the \imps. We deem it as the beginning of a research thread, and discuss more about its limitations and potential future directions in~\secref{sec:conclusion}.}
\pamout{Our choice for AS-level granularity is twofold: data availability and  scope. Specifically, at the AS-level there are several public datasets and rich information. For a finer granularity, one would need to conduct measurements and analyses to collect or infer the needed data, which can be very useful for several use cases, but is out of the scope (and space limitations) of this paper. Moreover, our study is the first aiming to comprehensively characterize the bias in the \imps. We deem it as the beginning of a research thread, and discuss more about its limitations and potential future directions in~\secref{sec:conclusion}.}
}

\myitem{"Vantage Points (\vps)".} Since we study bias at an AS-level, in the remainder, we will not differentiate between different probes in RIPE Atlas that are hosted in the same AS, or between different peers of RIPE RIS and \rv with the same ASN. And, for brevity, we will refer to the ASes that host RIPE Atlas probes or provide feeds to RIPE RIS / RouteViews as "vantage points" or \vps. 


\myitem{Dimension categories.} From the datasets we select \camr{all the characteristics that relate to the concept of bias, in order to make our analysis as general as possible. We end up to} \tma{a set of 22} characteristics that relate to the concept of bias and group them in the following categories: 


\begin{itemize}[leftmargin=*,nosep]
    \item \textbf{\textit{Location}}: RIR region; Country; Continent
    \item \textbf{\textit{Network size}}: Customer cone (\#ASNs,\#prefixes,\#addresses); AS hegemony; CTI "origin" and "top" indices
    \item \textbf{\textit{Topology}}: \#neighbors (total, peers, customers, providers)
    \item \textbf{\textit{IXP-related}}: \#IXPs; \#facilities; Peering policy
    \item \textbf{\textit{Network type}}: Net. type; Traffic ratio; Traffic volume; Scope; Personal ASN; ASDB classification (level 1 and 2)
\end{itemize}
\textit{Remark:} \camr{It is important to note that our methodology is generic and more characteristics can be included or grouped differently.} We only use these groups to facilitate the discussion in the paper (i.e., to refer to multiple dimensions under a single term), but we present detailed results for all dimensions.

\tma{Figure~\ref{fig:dataframe} depicts an example of the compiled dataset \camr{(which is also available in~\cite{ai4netmon-github})}.}



%% file: sections/analysis.tex
\begin{figure}
    \centering
    \includegraphics[width=0.9\linewidth]{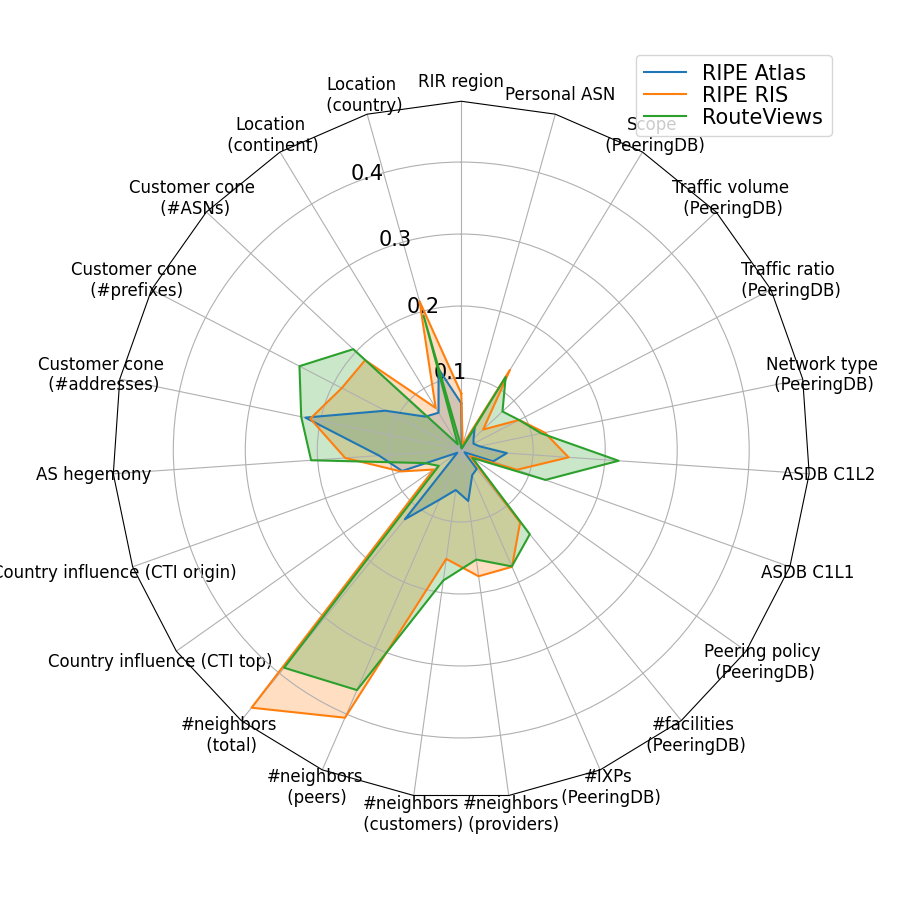}
    \caption{Radar plot depicting the bias score for RIPE Atlas (blue line/area), RIPE RIS (orange line), and RouteViews (green line) over the different dimensions (radius of the circle). Larger values of bias scores (i.e., far from the center) correspond to more bias.}
    \label{fig:radar-infrastructure-comparison}
\end{figure}

In this section, we study the biases in RIPE Atlas, RIPE RIS, and RouteViews.
Figure~\ref{fig:radar-infrastructure-comparison} shows a radar plot with bias scores for all dimensions. 
The colored lines---and their included area---correspond to the bias metric of a given \imp along a given dimension, e.g., the bias score for RIPE RIS (orange line) in the dimension “Location (country)” is 0.2. Larger bias scores (i.e., farther from the center) correspond to more bias,  e.g., in the dimension “Location (country)” RIPE RIS is more biased than RIPE Atlas (blue line). \pamout{Values closer to the center indicate lower bias.} 

\textit{Remark:} As knowing the entire distribution of a characteristic may help to better understand the bias along a certain dimension, we provide detailed distribution plots (i.e., similar to those in Fig.~\ref{fig:example-distribution-characteristics}) for all characteristics in \camr{the extensive documentation of our code and data~\cite{ai4netmon-github}}.

\begin{figure}
    \centering
    \includegraphics[width=0.95\linewidth]{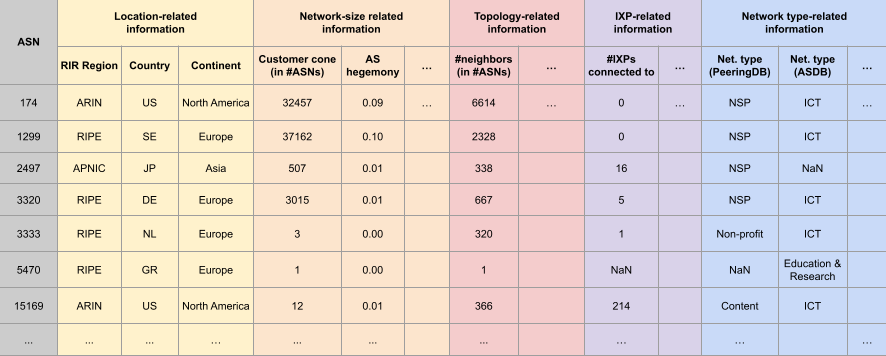}
    \caption{\tma{An example depicting the compiled dataset with characteristics (columns) of ASes (rows).}}
    \label{fig:dataframe}
\end{figure}


\myitem{Key findings.} Based on Fig.~\ref{fig:radar-infrastructure-comparison}, we can observe that:
\begin{itemize}[leftmargin=*, nosep]
    \item While the bias of \imps differs significantly by dimension, RIPE Atlas is substantially less biased than RIPE RIS and \rv along most dimensions.
    \item RIPE RIS and \rv have significant topological bias (e.g., number of neighbors/peers) as most of their collectors are deployed at IXPs, where ASes establish many (peering) connections~\cite{prehn2022peering}. 
    \item \rv and RIPE RIS are also quite biased in terms of network size (“Customer cone” dimensions) because they peer with many large ISPs. Having feeds from large ISPs may be desired for visibility, however, users still should be aware of it since it may lead to biased measurements.
    \item In most IXP-related and network type dimensions (that correspond to data mainly from PeeringDB), all platforms have relatively low bias; with an exception of RIPE RIS and \rv that are biased in terms of number of IXPs/facilities the \vps are connected to.
    \item There are small differences between RIPE RIS and \rv. RIPE RIS is more biased in terms of topology (number of neighbors, total and peers), whereas RouteViews is more biased in terms of network sizes (“Customer cone” and “AS hegemony” dimensions).
    \item We applied the KS-test for all platforms and dimensions. In almost all cases, the KS-test rejected the null hypothesis that the \imps vantage points follow the same distribution as the entire population of ASes. The only exceptions were the "Personal ASN" dimension for all \imps, and the "RIR region" and "Location (continent)" for \rv (where bias scores are less than 0.01).
\end{itemize}

\noindent\tma{\camr{Table}~\ref{tab:corr-dimensions} shows the correlation between the network characteristics for the entire population of ASes. The characteristics are grouped in the categories of \secref{sec:bias-dimensions}, and values correspond to averages among groups (i.e., values in the diagonal are not 1)\footnote{
\tma{Since our dataset consists of both numerical and categorical data, we use (i) the Pearson correlation coefficient for pairs of numerical features, (ii) the correlation ratio for pairs of a numerical and a categorical feature, and (iii) Cramer's V test for correlations between categorical features.}
}. As expected, dimensions in the same category are correlated. Also, topology dimensions are significantly correlated with IXP-related dimensions. Nevertheless, comparing with Fig.~\ref{fig:radar-infrastructure-comparison}, we see that correlated dimensions do not necessarily share similar bias scores. This highlights that a multi-dimensional bias exploration (Fig.~\ref{fig:radar-infrastructure-comparison}) can give more insights.}


\begin{table}[b]
    \centering
    \caption{\camr{Correlations between dimensions categories for the entire population of ASes.}}
    \label{tab:corr-dimensions}
    \begin{footnotesize}
        \begin{tabular}{r|ccccc}
       & Location  & Net. type & IXP-rel. & Topology & Net. size  \\
       \hline
       Location  & 0.99  & 0.15 & 0.15 & 0.24 & 0.21 \\
       Net. type  & & 0.31 & 0.33 & 0.26 & 0.17 \\
       IXP-rel.  & & & 0.75 & 0.48 & 0.26 \\
       Topology  & & & & 0.69& 0.38\\
       Net. size  & & & & & 0.40
        \end{tabular}
    \end{footnotesize}
\end{table}




\begin{figure*}[t]
    \centering
    \subfigure[RIS \& RV]{\includegraphics[width=0.3\linewidth]{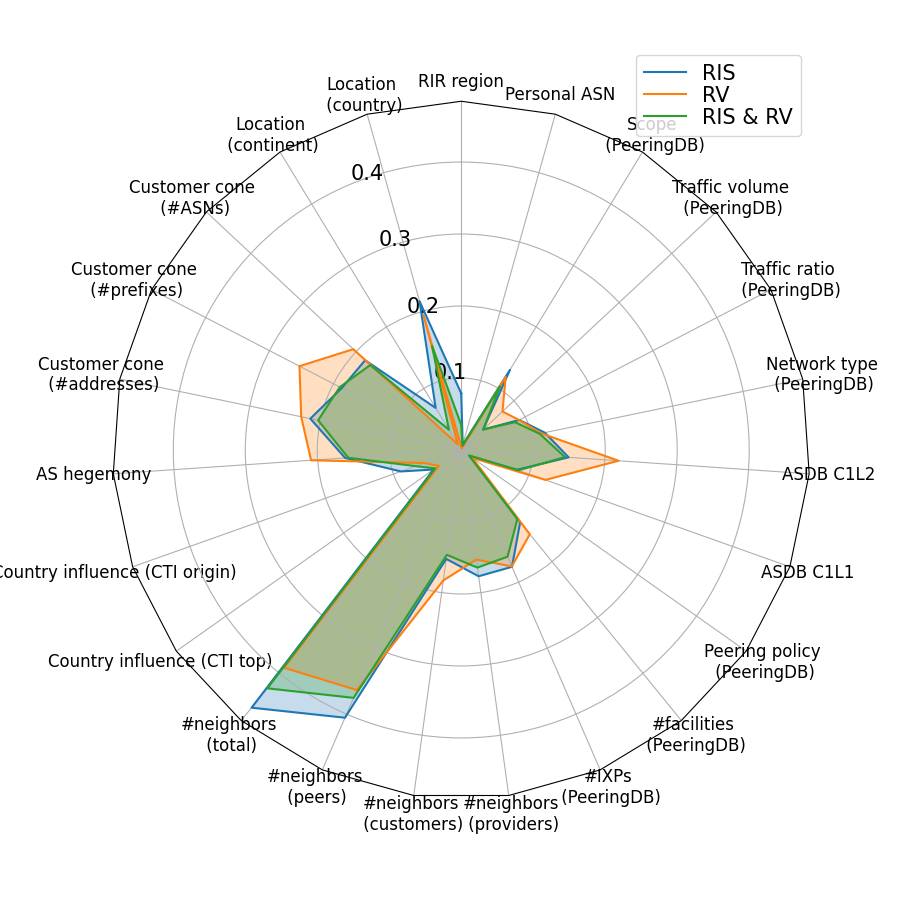}\label{fig:radar-infrastructure-special-cases-ris-and-rv}}
    \subfigure[Full feeds]{\includegraphics[width=0.3\linewidth]{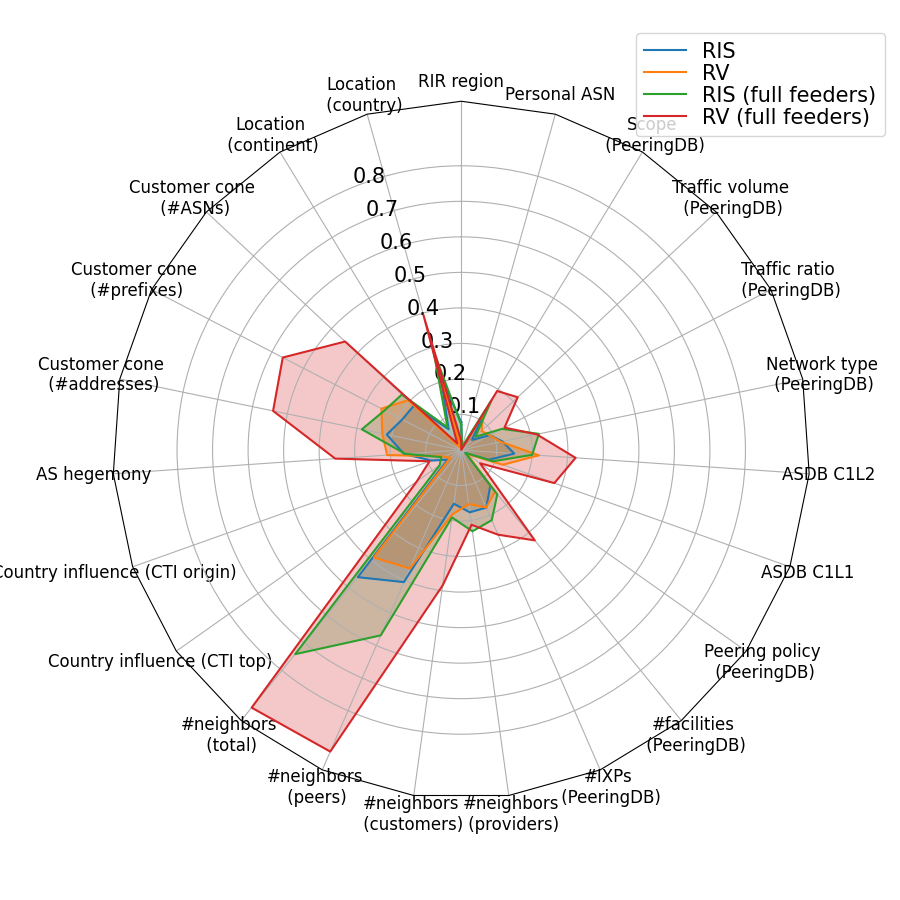}\label{fig:radar-infrastructure-special-cases-full-feeders}}
    \subfigure[Atlas: IPv4 vs IPv6]{\includegraphics[width=0.3\linewidth]{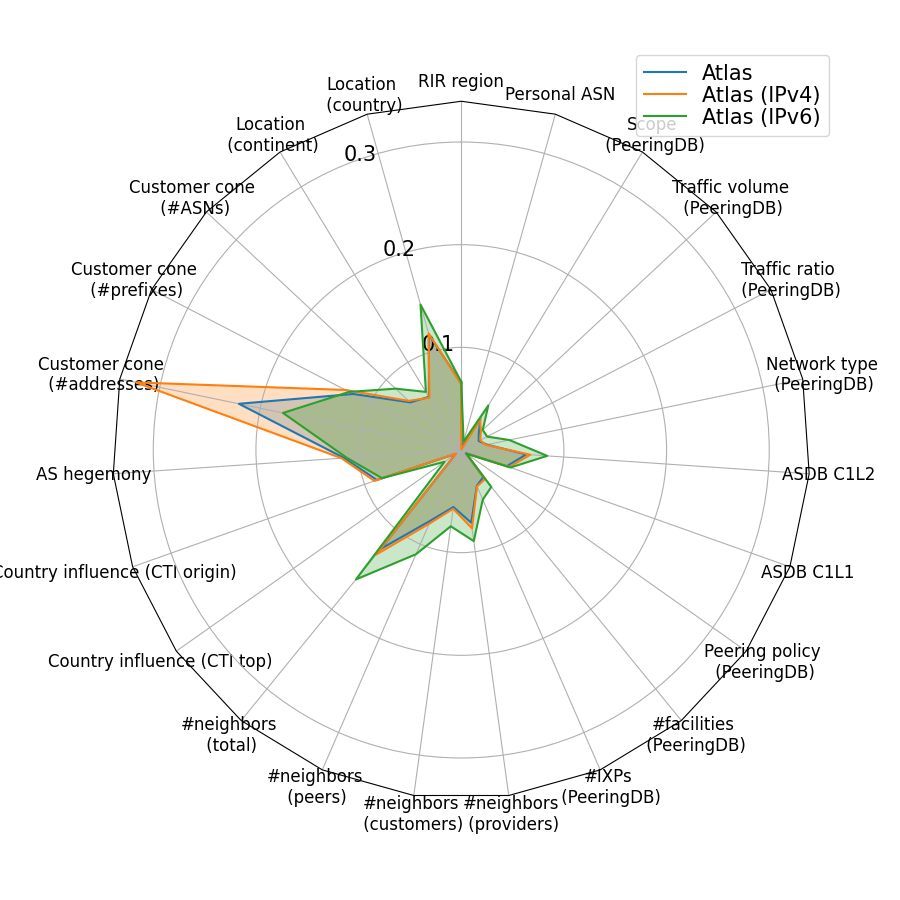}\label{fig:radar-infrastructure-special-cases-atlas-ipv}}
    \caption{Radar plot of bias score for the cases of (a) the combined RIPE RIS and \rv vantage points, (b) full feeds (vs all feeds) for RIPE RIS and \rv, and (c) IPv4 vs IPv6 RIPE Atlas probes. Note the different bias ranges in each plot.}
    \label{fig:radar-infrastructure-special-cases}
\end{figure*}

\vspace{\baselineskip}
Beyond this basic analysis, we conduct three similar analyses deepening our understanding of different \imp aspects. 

\myitem{Combining RIS and \rv.} Using data from both RIPE RIS and \rv is common (e.g., via CAIDA BGPStream~\cite{caidabgpstream}); hence, we analyze the combined bias in Fig.~\ref{fig:radar-infrastructure-special-cases-ris-and-rv}. When considering vantage points from both projects, the bias slightly decreases in most dimensions. Interestingly, there are some exceptions, e.g., number of neighbors (total and peers), where it would be preferable---in terms of bias---to use only feeds from \rv.

\myitem{Full vs. all feeds.} Only 240 and 70 peers of the RIPE RIS and \rv peers provide feeds for the entire routing table ("full feeds"), respectively. Figure~\ref{fig:radar-infrastructure-special-cases-full-feeders} compares the bias of only full feed peers against the entire \imps.  For RIPE RIS the increase in bias is small, whereas for \rv the set of full feeds is significantly more biased. In fact, while RIPE RIS is on average more biased than \rv, the opposite becomes true when considering only full feeds.

\myitem{IPv4 vs IPv6 vantage points.} Figure~\ref{fig:radar-infrastructure-special-cases-atlas-ipv} compares the set of ASes hosting IPv4, IPv6, and all RIPE Atlas probes \camr{(i.e., both IPv4 and IPv6)}. The set of networks hosting IPv6 probes is slightly more biased than networks hosting IPv4 probes in most dimensions. The only exception is the \#addresses in customer cone, which is mainly due to the differences in the IP space between the two versions. In RIPE RIS (not depicted in the plot), the differences between IPv4 and IPv6 peers is negligible. \camr{Due to the similarity in our analyses between IPv4 and IPv6 \vps, in the remainder we do not present separate results for each of these subgroups.}

\subsection{Analyzing improvement potential}\label{sec:improvement-potential}

Now that we have a basic understanding of the current biases in \imps, we want to compare the current state to a (hypothetical) case, where vantage points are randomly deployed among all types of networks, locations, etc. This comparison 
(i) provides a better understanding of the potentially avoidable \imp bias, and consequently (ii) reveals room for improvement (under practical limitations).

\myitem{Random sampling} from the entire population is an unbiased process\pamout{ (at least within the context of this paper)}. A sufficiently large random sample would lead to zero bias. Yet, small samples tend to be biased especially for characteristics with large variance. We treat the bias score that can be achieved via random sampling as a non-biased baseline.

Table~\ref{table:bias-infra-vs-random} compares the average bias over all dimensions\footnote{\pam{There are infinite options of combining bias scores of different dimensions. Here, we consider averaging as an intuitive choice, however, our framework supports other options as well (e.g., weighted average, or bias for subsets of dimensions).
}} of the \imps against that of a random sample with the same number of vantage points (e.g., in the case of RIPE RIS we consider random samples of size $|\mathvp|$=539). We repeat our random sampling 100 times and report the average bias. We observe that \textit{with the same number of \vps as in the current \imps, a random sample of ASes would have on average (almost) no bias}. 
This indicates that the ``limited'' number of vantage points is not the root cause of bias (which is mostly due to the deployment strategies; see \secref{sec:bias-examples-infra})
. 
\tma{In other words, \textit{we do not need more \vps, but more representative \vps}. This finding can be valuable for future extensions of \imps (e.g, selection of deployments in under-represented parts of the Internet) and improvement of measurement techniques (e.g., carefully selecting representative subsets of existing \vps); we identify these two aspects as key future research directions that can stem from our framework.}
\tmaout{In fact, we show later that adding few well-chosen \vps can drastically reduce the overall bias (\secref{sec:acquisition}) and that very low-biased \imp subsets can be selected via subsampling (\secref{sec:sampling-methods}).}

\begin{table}[b]
    \centering
    \caption{Bias of \imps vs. random sample of vantage points.}
    \begin{footnotesize}
    \begin{tabular}{l|cccc}
         {Platform} & {Atlas} & {RIS} & {RV} &{RIS \& RV}\\
         (\#vantage points) & {(3391)} & {(539)} & {(340)} & {(762)}\\ 
         \hline
         Platform bias &0.06 & 0.16 & 0.15 & 0.14\\
         Random sample bias &0.00 & 0.01 & 0.01 &0.01 
    \end{tabular}
    \end{footnotesize}
    \label{table:bias-infra-vs-random}
\end{table}

\myitem{Bias vs. number of vantage points.} 
While the current set of \vps is clearly not optimal in terms of bias, we wonder how bias changes when we only use a smaller random set of \vps (e.g., measurements with few Atlas probes due to rate/credit limits, or collecting feeds from a subset of route collectors peers due to the large volumes of data~\cite{alfroy2022mvp}). Figure~\ref{fig:bias-vs-sample-size-total} shows the average bias for different sample sizes drawn randomly from either the entire population of ASes (`all') or one of the three \imps. Lines correspond to averages over 100 sampling iterations, and errorbars indicate 95\% confidence intervals. For ease of comparison, dashed lines correspond to the bias values of using the entire infrastructure (i.e., the values in Table~\ref{table:bias-infra-vs-random}). We observe that: \one the bias decreases with the sample size (as expected), \two random sampling ("all") always has lower bias for the same number of \vps, \three even for very small samples, $\geq$20 \vps, random sampling ("all") has lower bias than the \textit{entire} sets of RIPE RIS and \rv \vps (dashed lines), while the same holds for RIPE Atlas for $\geq$40 \vps.

For a deeper inspection of the bias in smaller sets of \vps, Fig.~\ref{fig:bias-vs-sample-size-radar-rv} presents the bias of random samples of \rv \vps of sizes 10, 20, and 100 (similar results hold also for RIPE RIS and Atlas). We can see how the bias decreases in all dimensions for larger subset sizes. Yet, the change in bias is not the same in all dimensions, e.g., in network type dimensions the relative increase in bias for small subsets is much larger than in topology. 
\tma{This type of analyses can help to design measurements, for example, to select the number of \vps \camr{(e.g., see~\cite{reuter2018towards} or~\cite{sermpezis2021estimating})} based on the required level of bias per dimension.}



\subsection{Bias in common measurement practices}\label{sec:bias-common-measurements}
In this section, we briefly analyze the bias involved in common \vp selection methods that users follow in practice. 

\myitem{RIPE Atlas probe selection algorithm.} RIPE Atlas users can either select specific probes to use in their measurements or not specify them (which is the default choice; with parameters 10 probes from ``worldwide locations''\footnote{\url{https://atlas.ripe.net/docs/udm#probe-selection}}). In the latter case, RIPE Atlas has an automated algorithm to assign probes to a measurement, which prioritises probes with less load over more loaded probes, which makes the probe selection procedure not equivalent to true random sampling. 

\begin{figure}
    \centering
    \subfigure[\camr{Average bias vs. \#\vps}]{\includegraphics[width=0.44\linewidth]{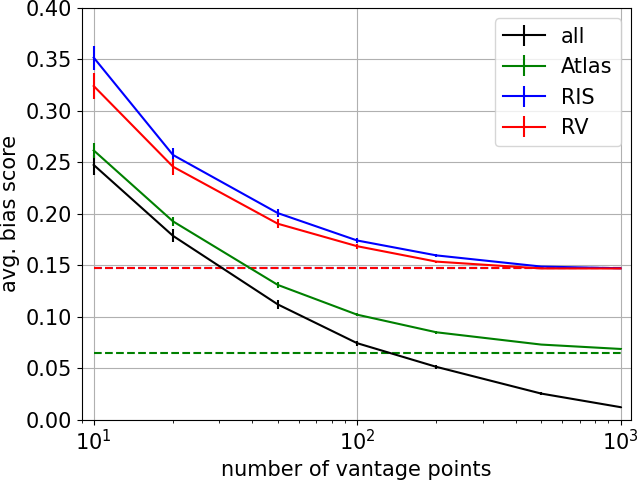}\label{fig:bias-vs-sample-size-total}}
    \subfigure[\camr{Bias of \rv subsets}]{\includegraphics[width=0.8\linewidth]{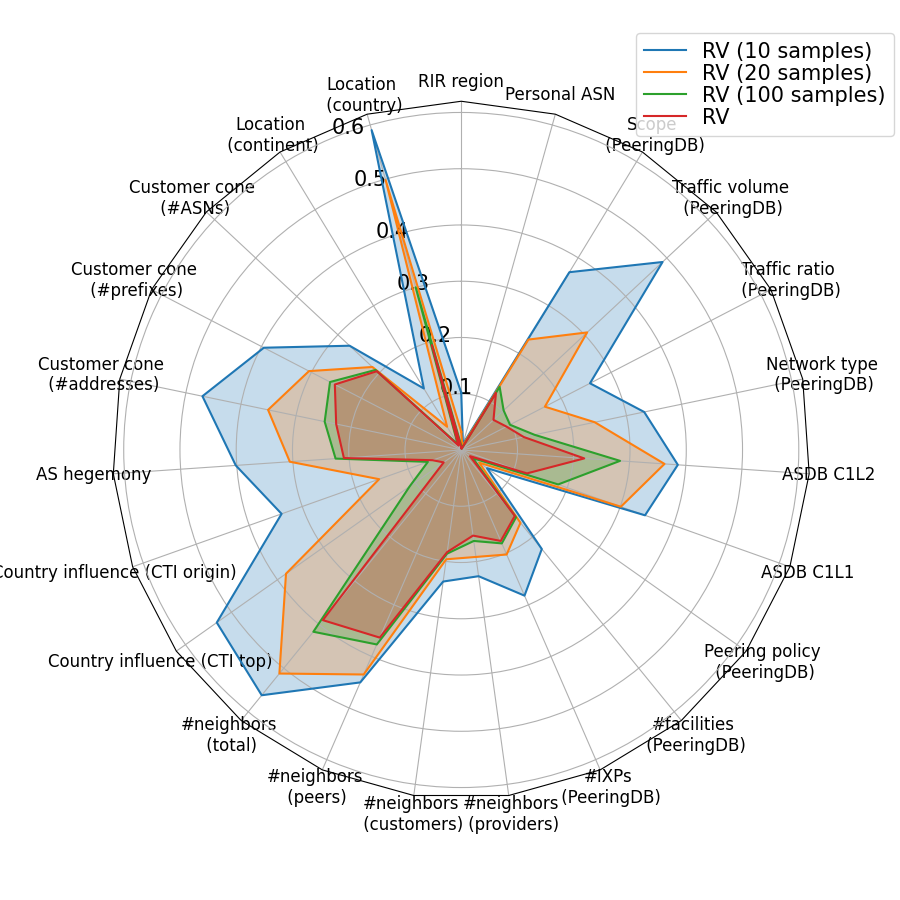}\label{fig:bias-vs-sample-size-radar-rv}}
    \caption{Bias vs. \#\vps: (a) Average bias (y-axis) of random samples from the entire population of ASes ("all") and from the \imp \vps vs. sample size (x-axis); (b) Bias of different sample sizes of \rv \vps in all dimensions.}
    \label{fig:bias-vs-sample-size}
\end{figure}

\begin{figure}
    \centering
    \subfigure[Atlas probe selection]{\includegraphics[width=0.44\linewidth]{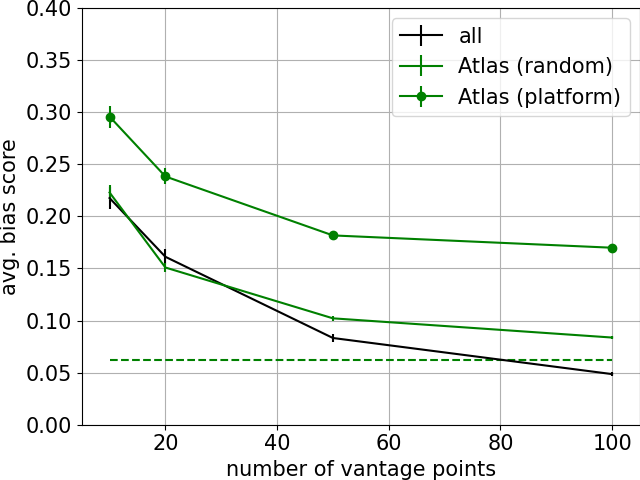}\label{fig:bias-vs-sample-size-total-real-Atlas}}
    \hspace{0.08\linewidth}
    \subfigure[Bias vs. nb of \vps per RC]{\includegraphics[width=0.44\linewidth]{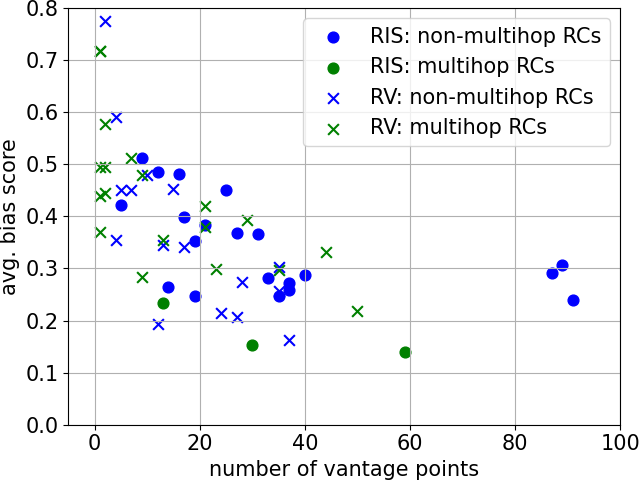}\label{fig:bias-vs-nb_peering-asns}}
    \caption{(a) Average bias score vs. sample size for the RIPE Atlas probe selection algorithm vs. random sampling. (b) Scatter plot of average bias (y-axis) vs. number of \vps (x-axis) per route collector of RIPE RIS and \rv.}
    \label{fig:bias-common-practices}
\end{figure}

In Fig.~\ref{fig:bias-vs-sample-size-total-real-Atlas} we study how the RIPE Atlas selection algorithm, "Atlas (platform)", performs compared to random sampling from either all RIPE Atlas probes, "Atlas (random)", or from all ASes ("all"); the values for these latter cases are the same as in Fig.~\ref{fig:bias-vs-sample-size-total}). We considered the sets of probes that the RIPE Atlas platform returned when we initiated measurements with parameters \texttt{type="area"} and \texttt{value="WW"}. Lines correspond to averages over 100 sampling iterations, and errorbars indicate 95\% confidence intervals. 
We observe that \textit{when using the RIPE Atlas algorithm for selecting probes, "Atlas (platform)", then the bias is significantly higher compared to randomly selecting probes, "Atlas (random)"}. In fact, the bias is almost two times higher. This indicates that even with the existing infrastructure, users could decrease bias by 50\% by not depending on the built-in probe selection process, but select random probes themselves.

\myitem{Feeds from a single route collector (RC)} may be used in cases that there are processing limitations (e.g., in terms of real-timeness or storage) due to the large volume of data, see~\cite{green2018leveraging,ariemma2021long,alfroy2022mvp}. Figure~\ref{fig:bias-vs-nb_peering-asns} presents the average bias score per RC (i.e., the bias of the set of \vps that peer to a RC) in relation to its number of \vps. Overall, there is a clear (negative) correlation between the number of \vps and the bias score of a RC. Nevertheless, the size of a route collector does not predict its bias as \one the three RCs of RIPE RIS (rrc01, rrc03, rrc12) that are significantly larger ($>$80 members) than the rest of RCs, are not less biased (in fact, there are several smaller RCs with lower bias) and \two there are several medium-size RCs (and even some with only 10-20 \vps) that have relatively low bias. For RIPE RIS, the three multihop RCs (rrc00, rrc24, rrc25) are less biased than most of the non-multihop RCs (which are deployed at IXPs). \tma{For further analyses, we provide through our API and online tools (\camr{\secref{sec:portal};~\cite{ai4netmon-website}}) the detailed bias scores and radar-plots for each RC.} 


\camr{
\textit{\myitem{Summary of main takeaways:} (i) RIPE RIS and \rv are substantially more biased than RIPE Atlas; and their \vps are significantly more biased towards networks at IXPs (with many peering links) and larger networks. (ii) Considering only full-feeds further intensifies the bias of control-plane \imps, while differences between IPv4 and IPv6 \vps are less important in all \imps. (iii) If \imps would choose \vps entirely random, their current set of \vps would be very close to an ideal sample; this indicates that bias is not due to the size of \imps, but mainly due to \vp deployment strategies. (iv) Common practices to limit the number of \vps yield higher bias than simple random samples from \imps.} 
}

%% file: sections/portal.tex
To facilitate users and further research and analyses, we provide data, code, and tools to calculate and visualize the bias in a set of networks\camr{~\cite{ai4netmon-website, ai4netmon-github}}. 

\myitem{Data.} The data we aggregated from different sources are provided as a table with rows corresponding to ASNs and columns to network characteristics (see~\secref{sec:bias-dimensions})~\cite{ai4netmon-github}. 

\myitem{Code.} We open-source the code for calculating the bias~\cite{ai4netmon-github}. The methods receive as input (i) the data table, (ii) a set of ASNs that are considered the ``population'' $\mathcal{N}$, (iii) a subset $\mathvp$ of the population, whose bias we are interested in, (iv) the set of characteristics that will be taken into account. \tma{This enables users to apply the framework in a generic way; for example:
\begin{itemize}[leftmargin=*,nosep]
    \item use other datasets than the ones we compiled in \secref{sec:bias-dimensions}
    \item perform a bias analysis with respect to a given region (e.g., \camr{setting as the ``population'' $\mathcal{N}$ only the ASNs in the RIPE region, instead of all the ASes we considered in this paper})
    \item explore the bias of a custom set of \vps $\mathcal{V}$ (e.g., a set of Atlas probes, or a set of route collector peers) 
    \item consider only a subset of bias dimensions
\end{itemize}
}

\myitem{Open API.} To further facilitate access to data and methods, we provide an API~\cite{ai4netmon-website} that \tma{provides (up-to-date) bias scores of the \imps we analyzed, other \imps (e.g., bpg.tools, CAIDA Periscope), individual route collectors (see \secref{sec:bias-common-measurements}), or any custom set of \vps (or, ASNs in general) requested by~a~user.}



\myitem{Web portal.} We provide a set of online tools for interactive visualizations of the bias data, namely, radar plots (as in Fig.~\ref{fig:radar-infrastructure-comparison}) and the detailed distributions per characteristic (CDF plots or histograms
) for all platforms~\cite{ai4netmon-website}. 


%% file: sections/related.tex
\myitem{Topological bias of route collectors.} When analyzing the Internet's topology, route collectors often miss many interconnections of CDNs~\cite{arnold2020cloud}, at IXPs~\cite{ager2012anatomy,prehn2022peering,giotsas2013inferring}, or due to complex routing setups~\cite{oliveira2008search}. While it is hard to remove these biases, many works tried to understand the importance of certain biases for their work by analyzing how their results would change when using only subsets of the available infrastructure, e.g.,~\cite{luckie2013relationships,jin2020toposcope,sermpezis2021estimating,marcos2020path},~\camr{\cite{reuter2018towards}}. \camr{While we do not focus on the effect of biases in measurements, our framework enables to easily quantify biases, e.g., in these subsets of the infrastructure, and thus provide further insights on how it affects or correlates with the resulting measurement biases.}

\myitem{Biases and use cases.} While Roughan \etal argued that route collectors are biased towards larger core networks and IXPs~\cite{roughan201110}, Chung \etal ~\cite{chung2019rpki} saw no substantial differences when comparing their view on the longitudinal deployment of route origin validation with that of Akamai gathered from an order of magnitude more monitors\footnote{The study only analyzed prefix-origin pairs that were visible by the route collectors. It remains unclear whether this result would change when also considering Akamai's privately received BGP announcements.}. \camr{On the other hand,~\cite{prehn2021biased} shows that \imp data can lead to significant geographical and topological biases in AS relationships inference.} This highlights that biases might be use-case dependant---a fact further supported by the work of Cittadini \etal from 2014 which showed that route collectors have different biases for topology analysis and iBGP policy inference~\cite{cittadini2014quality}. In 2009,~\cite{heidemann2009uses} argued that Internet measurements, in general, are biased by various (sometimes unknown) factors such as traffic volume, user populations, or topology, which is further supported by a series of exemplary experiments conducted by Bush \etal~\cite{bush2009internet}.

\myitem{Bias in RIPE Atlas.} In 2015, Bajpai \etal showed that the distribution of Atlas probes to ASes is heavy-tailed and also analyzed the network type distribution of probe hosting ASes (without comparing it to the overall type distribution)~\cite{bajpai2015lessons}. A later study by Bajpai \etal in 2017 further found that 91\perc of RIPE Atlas probes are located in the RIPE and ARIN region and that the number of probes is not representative for the number of Internet users in countries such as Japan~\cite{bajpai2017vantage}. \camr{These types of bias explorations are facilitated (and extended to more bias dimensions) by the proposed framework.}

\tmaout{
\myitem{Extending current platforms.}
In 2008, Roughan \etal model the topology discovery problem as an extension of the simple capture-recapture model that is frequently used in biological research to estimate the population of a species via $K$ random yet comparable samples. They argue that their model can also be used to estimate beneficial route collector peers and they estimate that significant global link coverage can be reached with fewer than 700 peers~\cite{roughan2008bigfoot}. 

In 2012, Gregori \etal observed that customer ASNs see most---if not all---of their provider's routing information. Based on this observation they introduced a metric named p2c-distance that counts the number of transit relationships an update has to travel between two ASNs. Their approach defines the optimal route collector set as the minimal number of ASes needed such that each AS in the entire Internet has at most a p2c-distance of $N$ (they practically used 2) to at least one feeder ASN. They solve a slightly modified minimum set cover problem to compute this set of ASNs and find that multi-homed stub ASNs are most valuable as new collector feeds~\cite{gregori2012incompleteness}.

In 2022, Leyba \etal analyzed AS topology from a probabilistic standpoint. They accumulated different observations over several time periods, and assigned to each link a probability of existence. Their model show the most uncertainty for links in Israel, Egypt, Georgia, Bulgaria, and Iceland, suggesting that connecting ASNs in these countries can be beneficial for topology discovery~\cite{leyba2022cutting}.
}

\myitem{\tma{Similarity of \vps.}} \tma{Two recent studies~\cite{appelmetis,alfroy2022mvp} considered similarity of RIPE Atlas and RIPE RIS \vps, respectively, for subsampling \vps and thus avoiding redundant (i.e., over-represented) information. \cite{appelmetis} calculates a similarity matrix between Atlas probes based on measurements, and proposes a method to select subsets of probes that are dissimilar. Similarly, \cite{alfroy2022mvp} calculates \vps similarities based on topological characteristics, and applies a clustering algorithm to select a set of dissimilar of \vps aiming to achieve a good tradeoff between volume of information (i.e., less \vps) and observability of the AS topology. We consider these works complementary to ours; a difference is that~\cite{appelmetis,alfroy2022mvp} are measurement-dependent, however, investigating relation between bias and \vp similarities can lead to more efficient subsampling methods.}

\tmaout{
\myitem{Subsampling \imps.} \pam{Two very recent studies~\cite{appelmetis,alfroy2022mvp} considered subsampling of RIPE Atlas and RIPE RIS vantage points, respectively. \cite{appelmetis} calculates a similarity matrix between Atlas probes based on measurements, and proposes a method to select subsets of probes that are dissimilar. Similarly, \cite{alfroy2022mvp} calculates \vps similarities based on topological characteristics, and applies a clustering algorithm to select a set of dissimilar of \vps aiming to achieve a good tradeoff between volume of information (i.e., less \vps) and observability of the AS topology.}
}

%% file: sections/conclusion.tex

This work aims to be the first effort for a systematic and comprehensive characterization of bias in \imps, by providing a framework to quantify bias (metrics, data, code, etc.) and an analysis of popular \imps. Being aware about the existence of bias and its "flavors" (e.g., how much and at what dimensions) can help the users of \imps to carefully interpret the results of their measurements, and avoid pitfalls or wrong generalizations that may appear due to the bias. 

\tma{Before our work, significant biases in IMP vantage point placements have been documented by experience papers from well-established scientists (e.g., ~\cite{roughan201110,bush2009internet}) or via a few dedicated analyses~\cite{holterbach2015quantifying,bajpai2015lessons,bajpai2017vantage}. Besides reproducing their original findings, the framework we introduced drastically facilitates finding new biases among diverse dimensions and tracking of the evolution of these biases over time. }

Moreover, our findings and tools (data, code, API) can further help users to fine-tune their measurements (e.g., select a set of vantage points), and provide useful insights to \imp operators for extending their platforms. We see several promising messages in our results towards these directions.

\pam{We deem our work as an initial (but, necessary) step towards a complete understanding of bias in \imps and its impact on user measurements. There are many research directions and improvements that would need a more extensive investigation and can be addressed in future work.} 

\pam{In the following we provide a critical discussion for some of these directions, in relation to our work:}

\myitem{AS-level granularity.} \pam{We conducted our analysis at an AS-level, because the majority of data sources provide data at this granularity. It is straightforward to generalize our framework to a more fine-grained level (all methods, metrics, etc., directly apply). For example, if we have available data per prefix\footnote{Some ASes consist of many---sometimes globally distributed---routers that make independent decisions, which can be captured at a prefix level.}, then we can consider as our "population" all the routed prefixes, and as "sample population" the prefixes that contain the IP addresses of the RIPE RIS / \rv peers or the RIPE Atlas probes. Our methods would then simply take as input a table \tma{as in Fig.~
\ref{fig:dataframe}} with rows the prefixes (instead of ASes) and columns the prefix characteristics (instead of AS characteristics).}

\pam{Several use cases could benefit from such a more fine-grained granularity. However, the challenging part is the data availability. To extract even a single characteristic at this granularity, we may need extensive measurements and analyses. For example, a custom method is needed to infer per-prefix locations~\cite{winter2019geo}, while to infer customer cones per-prefix could lead to incomplete data since aggregating measurements from \vps in different prefixes would not be possible.}

\myitem{Dimensions of bias (per use case).} Not all dimensions of bias may be relevant to a measurement study. For example, any bias in the "peering policy" dimension may not affect latency measurements (this is just a conjecture)\pam{, whereas it is probable to affect BGP hijacking detection measurements}. \tma{In another example,~\cite{sermpezis2021estimating} has shown that estimating the impact of a hijack with RIPE RIS and \rv leads to a 10\% higher error than custom measurements from random (i.e., unbiased) ASes; this error due to bias may be lower/higher in a different use case though.} Identifying which dimensions are important per use case, could improve our understanding of bias and its role. However, this requires a per case analysis, since there are many different measurement use cases with a wide range of scopes and objectives. \tma{Our tools (\secref{sec:portal}) enable to exclude dimensions, thus, covering as many use cases as possible.}

\tmaout{
Our framework allows to account for different use cases, by considering bias differently per dimension. Specifically, while in our paper we decrease the \textit{average} bias score along all dimensions (\secref{sec:sampling-methods} and \secref{sec:extra-infrastructure}), one can define different aggregation ways (i.e., other than averaging). More formally, if $B_{i}(\mathcal{S})$ is the bias score of a set $\mathcal{S}$ along a dimension $i$, then the overall bias score can be any function $f$ of the individual scores: $B(\mathcal{S}) = f\left( B_{1}(\mathcal{S}), B_{2}(\mathcal{S}), ..., B_{K}(\mathcal{S}) \right)$. Some examples could be: (i) a weighted average, where the importance of each dimension would be captured by a weight $w_{i}$ (which can be tuned per use case), or (ii) the "stricter" case of $\max_{i}B_{i}(\mathcal{S})$ that captures the "worst case" of bias, or (iii) $B(\mathcal{S}) = 1 - \prod_{i=1,2,..., K} \left(1-B_{i}(\mathcal{S})\right)$ that aims to achieve a "balance" among all dimensions.
}

\tmaout{
\myitem{Impact of bias.} In the CDN use case we considered, bias is responsible for a 6\% in the latency estimation error (see randomly selected vs. low-bias sets of Atlas probes; Table~\ref{table:average-error-cdn}). Another study~\cite{sermpezis2021estimating}, has shown that estimating the impact of a hijack with RIPE RIS leads to a 10\% higher error than custom measurements to random ASes. Knowing the impact of bias in a use case can help us build new methods (or even inform us to not focus on it, if the impact is small). However, as in our above discussion, this would need a per case analysis.
}

\myitem{Accuracy, completeness, and bias in ground truth data.}
\pam{The input to our framework (i.e, the AS characteristics) is from public datasets. And, some of them are known to suffer from inaccuracies (e.g., country information per ASNs), incompleteness (e.g., only 25\% of ASNs have records in PeeringDB), or even biases (e.g., data inferred based on measurements from the existing –biased– platforms, such as customer cones, topology, etc.). Improving the datasets would be beneficial, in general, and for the quantification of bias, in particular, since they could reveal further insights\footnote{The main insights of this paper are not expected to deviate significantly, since we have not identified any counterintuitive findings in our analysis.}; nevertheless, this is an orthogonal task.} \tma{As already discussed, changing the datasets does not change our framework, but only its input.}

\tmaout{
\myitem{\pam{Generalization of our framework: beyond \imps}.}
\pam{The population (i.e., set $\mathcal{N}$) does not necessarily be the entire population of ASes; depending on the use case, it can be the set of clients of a network, or a set of networks with a given characteristic (e.g., all ASes in a continent, or non-stub ASes), etc. Similarly, the selection of the subset $\mathvp$ may not be limited to \imps; e.g., it can any arbitrary set of networks that can be measured. In this way, our framework can be used to quantify biases in setups other than \imps; one just needs to change the input sets ($\mathcal{N}$ and $\mathvp$) in our methods, while the data remain the same.}
}

\tmaout{
\myitem{Community Contribution.}
Before our work, significant biases in IMP vantage point placements have been documented by experience papers from well-established scientists (e.g., ~\cite{roughan201110,bush2009internet}) or via a few dedicated analyses~\cite{holterbach2015quantifying,bajpai2015lessons,bajpai2017vantage}. Besides reproducing their original findings, the framework we introduced drastically facilitates finding new biases among diverse dimensions and can enable the tracking of the evolution of these biases over time. We demonstrated, e.g., that while IXP peering oriented networks are over-represented in RIPE RIS, their peering policies are representative of the Internet’s peering ecosystem\footnote{as captured by PeeringDB.}. Our framework further provides the tools needed to counteract the impact of bias when extending IMP infrastructures or choosing an unbiased set of vantage points for measurements.
}


%% file: appendix.tex
\section{Survey on Internet measurements and bias}\label{appendix:survey}
\input{sections/appendix_survey}

\section{Comparison of difference bias metrics}\label{appendix:bias-metrics}
\input{sections/appendix_metrics}

\tmaout{
\section{Different approaches for mapping complexity scores to ASNs}\label{appendix:questionnaire-merging}
\input{sections/appendix_score_merging}
}


%% file: sections/appendix_survey.tex
We conducted an anonymous survey on Internet measurements and bias in \imps. 
In this paper, we only provide a pointer to a single high-level finding of the survey in \secref{sec:bias-examples-infra}, and we do not rely any of the content of the paper on it. For completeness, we provide a short description of the survey: 


\myitem{Questions.} We have asked participants to indicate measurement use cases, and the insights they aim to get from measurement data. We ask them what measurement types they use (control and/or data plane), what information they collect from them (e.g., latencies, BGP paths), what \imps they use, and what is their scope (e.g., if they target small or large geographic areas, or network types such as ISPs, CDNs, etc.).

We ask them the question (to whose answers we refer in this paper) "Is there any kind of bias in the measurement data collected for this use case?", giving them three possible answers to select from "Yes / Probably yes", "No / Probably not", "I don't know".

We also ask them if they believe that there are location / network-type biases, and how useful they would find if we could provide them with analyzes/tools that would show and mitigate bias. 

\myitem{Responses (relevant to this paper).} We have received responses from 50 participants, both network engineers/operators ($\sim$75\%) and researchers ($\sim$25\%). More than $\sim$80\% of the participants said that they are experienced users of \imps. 70\% uses RIPE Atlas in their measurements, and around 50\% use RIPE RIS and/or \rv. 

26 participants (52\%) replied "Yes / Probably yes" in the question about existence of bias, 14 (28\%) replied "No / Probably not", and 10 (20\%) replied "I don't know". 

%% file: sections/appendix_metrics.tex

\myitem{Bias metrics.} There are several metrics to quantify the difference between two distributions (i.e., the "bias" in our context). 
\begin{itemize}
    \item Kullback–Leibler (KL) divergence; see \secref{sec:bias-definition}
    \item Total Variation (TV) distance: $B_{TV} =  \sum_{i=1}^{K}|p_{i}-q_{i}|$
    \item Max distance: $B_{max} =  \max_{i=1}^{K}|p_{i}-q_{i}|$
\end{itemize}

The main difference between KL-divergence and TV distance metrics, is that the former is more sensitive to changes in characteristics of lower probabilities $p_{i}$~\cite{steck2018calibrated}. For example, let $P=[0.6,0.2,0.2]$ and two distributions $Q^{A}=[0.7,0.1,0.2]$ and $Q^{B}=[0.6,0.1,0.3]$ that differ by $\pm 0.1$ compared to $P$. While for the total variation it holds that $B_{TV}(P,Q^{A})=B_{TV}(P,Q^{B})$, for the KL-divergence it holds $B_{KL}(P,Q^{A}) < B_{KL}(P,Q^{B})$, because the $+0.1$  was at a characteristic with a lower probability in $Q^{B}$.

The main difference between the \textit{Max} distance and the other metrics, is that the former accounts for the "worst case" (i.e., max deviation between two distributions), whereas the latter calculate distances over the entire distribution.

\myitem{Bias in \imps for each metric.} In Fig.~\ref{fig:bias-vs-metrics} we present the radar plot depicting the bias for the three bias metrics. While the actual values differ for different metrics, the qualitative findings (e.g., which infrastructure set is more biased) remain the same for the majority of dimensions.

%% file: sections/appendix_score_merging.tex
By the end of our labeling process, most ASes have multiple labels. Given that we extracted the min, mean, and max answers for each label from the questionnaire and could take the min, max, or some merged value across all labels that are assigned to an ASN, we end up with nine different variations of how we could collapse the label set of an AS to a single ease-of-acquisition value. Please note that, instead of simply averaging them, we decided to merge values as follows: \one if there is -3 label, set the final value to -3, \two if there is a +3 label, set the final value to +3, and \three take the mean across all labels; this approach ensures that our final value always complies with ``guaranteed not`` and ``guaranteed.`` Before analysing the results and deciding for a collapsing policy, we divided all values by 3 to normalize them. 

\begin{figure}
\includegraphics[width=0.7\linewidth]{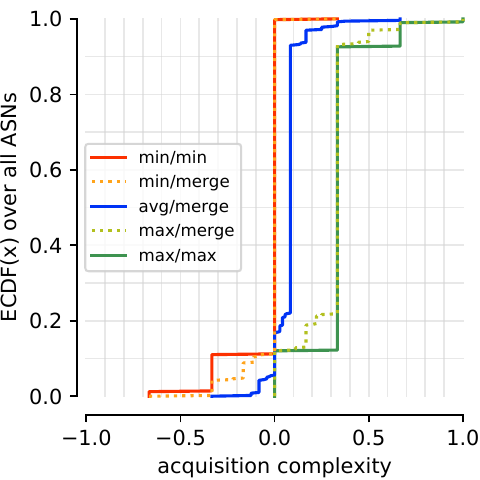}
\caption{Distribution of Acquisition complexity using different aggregation approaches.}\label{fig:acq:weights}
\end{figure}

Figure~\ref{fig:acq:weights} shows the ECDF of ease-of-acquisition values across all ASes for five of the before-mentioned nine collapsing approaches (the remaining four curves lie between the min/min and max/max curve and are hidden to preserve the clarity of the plot).  We observe that all curves show similar tendencies: Approximately 80\perc of ASes are equally hard to peer with, while the remaining 20\perc split almost equally between ASes that are substantially harder/easier to peer with. As the insides from the min/min and max/max curve are properly reflected by it, we decides to consider the avg/merge collapsing approach for the analyses in the paper.

